\documentclass[preprint,showkeys,superscriptaddress,amsmath,amssymb]{revtex4-1}

\usepackage{times}
\usepackage{graphicx}
\usepackage{dcolumn}
\usepackage{amsmath}
\usepackage{color}
\usepackage{caption,setspace}
\usepackage{amssymb}
\usepackage[version=3]{mhchem}

\begin{document}

\title{Efficient Dielectric Metasurface Collimating Lenses for Mid-Infrared Quantum Cascade Lasers}

\author{Amir Arbabi}
\affiliation{T. J. Watson Laboratory of Applied Physics, California Institute of Technology, 1200 E. California Blvd., Pasadena, CA 91125, USA}
\author{Ryan M. Briggs}
\affiliation{Jet Propulsion Laboratory, California Institute of Technology, Pasadena, CA 91109, USA}
\author{Yu Horie}
\affiliation{T. J. Watson Laboratory of Applied Physics, California Institute of Technology, 1200 E. California Blvd., Pasadena, CA 91125, USA}
\author{Mahmood Bagheri}
\affiliation{Jet Propulsion Laboratory, California Institute of Technology, Pasadena, CA 91109, USA}
\author{Andrei Faraon}
\email{faraon@caltech.edu}
\affiliation{T. J. Watson Laboratory of Applied Physics, California Institute of Technology, 1200 E. California Blvd., Pasadena, CA 91125, USA}

\begin{abstract}
Light emitted from single-mode semiconductor lasers generally has large divergence angles, and high numerical aperture lenses are required for beam collimation. Visible and near infrared lasers are collimated using aspheric glass or plastic lenses, yet collimation of mid-infrared quantum cascade lasers typically requires more costly aspheric lenses made of germanium, chalcogenide compounds, or other infrared-transparent materials.  Here we report mid-infrared dielectric metasurface flat lenses that efficiently collimate the output beam of single-mode quantum cascade lasers. The metasurface lenses are composed of amorphous silicon posts on a flat sapphire substrate and can be fabricated at low cost using a single step conventional UV binary lithography. Mid-infrared radiation from a 4.8 $\mu$m distributed-feedback quantum cascade laser is collimated using a  polarization insensitive metasurface lens with 0.86 numerical aperture and 79\% transmission efficiency. The collimated beam has a half divergence angle of 0.36$^\circ$ and beam quality factor of $M^2$=1.02.
\end{abstract}


\maketitle

Quantum cascade lasers (QCLs) are compact and efficient sources of coherent mid-infrared (mid-IR) radiation. Their small size, high spectral purity, and relatively low power consumption enable realization of compact mid-IR spectroscopy systems with applications in chemical sensing for environmental and medical monitoring~\cite{Kosterev2002,Bakhirkin2005,Kosterev2007,Briggs2014,Jouy2014}.
Due to the wavelength-scale emission aperture of single-mode QCLs, their output beams have large divergence angles. High numerical aperture (NA) aspheric lenses made of germanium or chalcogenide compounds, such as zinc selenide and black diamond (\ce{Ge_{28}Sb_{12}Se_{60}}), are conventionally used for collimation and tight focusing of mid-IR radiation. Several techniques have been previously proposed to reduce the beam divergence and to lower the cost of the collimation optics. Integration of directional antennas with the laser~\cite{Amanti2007,Danylov2007,DeglInnocenti2014}, and modification of the laser emission aperture~\cite{Yu2007,Yu2008,Yu2008a,Amanti2009} are shown to increase the effective area of the emission aperture and reduce beam divergence angle. Wafer scale fabrication of low NA multi-level mid-IR Fresnel lenses have been shown using multiple lithography and etching steps~\cite{Fonollosa2008}. Refractive mid-IR silicon lenses have been also demonstrated using photoresist reflow and dry etching processes~\cite{Logean2012}. 
Here we propose and experimentally demonstrate high NA diffractive mid-IR lenses based on dielectric metasurfaces. Diffractive elements based on metallic and dielectric metasurfaces have recently attracted significant attention~\cite{Kildishev2013a,Yu2014}. These metasurfaces are composed of two-dimensional arrays of subwavelength scatterers that modify the wavefront, polarization, or amplitude of light. They accurately realize lithographically defined phase, polarization, or amplitude profiles, and are fabricated using standard micro- and nano-fabrication techniques with potential for low cost wafer scale production. Visible and near-infrared dielectric metasurface lenses and focusing mirrors have been recently demonstrated~\cite{Fattal2010,Lu2010,Klemm2013,Lin2014a,Arbabi2014,Vo2014,Arbabi2015,Arbabi2015a}. For efficient operation in mid-IR, the metasurface lenses should be fabricated using scatterers and substrates with low absorption. We use hydrogenated amorphous silicon (a-Si) scatterers since a-Si has high refractive index and low loss in mid-IR~\cite{Palik1998}. The high refractive index contrast between the scatterers and their surroundings is essential for realization of high NA metasurface lenses~\cite{Arbabi2015}. Materials with low refractive index and low mid-IR absorption such as zinc selenide, calcium fluoride, magnesium fluoride, and sapphire can be used as the substrate for the metasurface lens. We chose sapphire for its wider availability, lower cost, and low absorption at wavelengths below 5 $\mu$m.

\begin{figure}
\centering
\includegraphics[width=0.85\columnwidth]{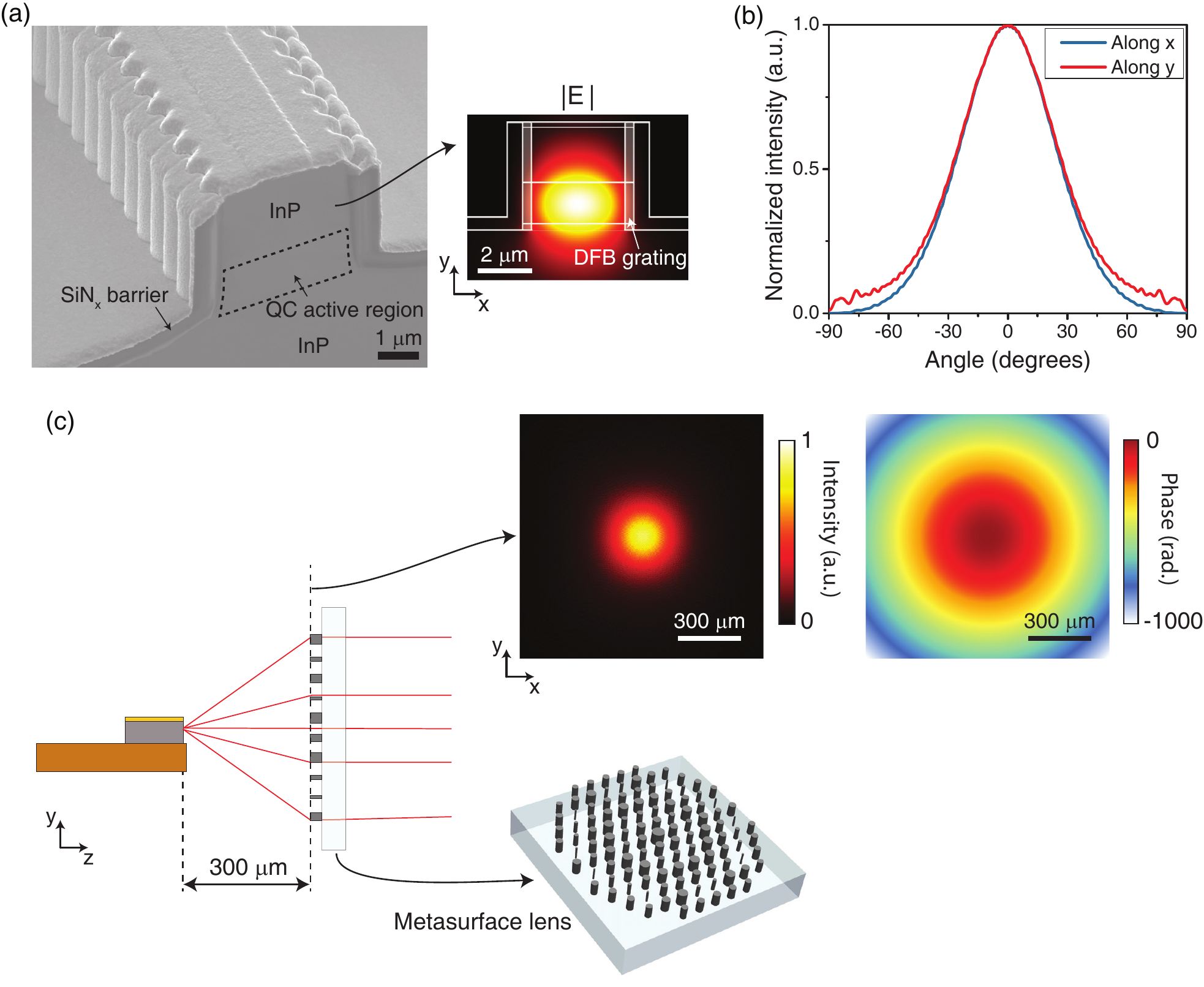}
\caption{(a) Scanning electron micrograph of the laser ridge waveguide and facet of a distributed-feedback QCL. The inset shows the simulated amplitude of the electric field of the lasing mode. (b) Simulated far-field emission pattern of the QCL shown in (a). (c) Schematic illustration of the QCL collimation using a metasurface lens. The inset shows the laser intensity and phase distributions at the lens plane.}
\label{fig:QCL}
\end{figure}

A scanning electron micrograph of the mid-IR QCL we use in this study is shown in Fig.~\ref{fig:QCL}(a). The QCL has a ridge width of 4 $\mu$m, is single mode, and emits dominantly transverse magnetic (y polarized) mid-IR radiation with wavelength of approximately 4.8 $\mu$m.  The sinusoidal corrugation of the ridge forms a first order distributed-feedback (DFB) grating that leads to single longitudinal mode operation of the laser. More details on the design, fabrication, and characterization of the laser are found in~\cite{Briggs2014a}. The amplitude of the electric field of the lasing mode is shown in the inset of Fig.~\ref{fig:QCL}(a). The mode has subwavelength dimensions along both the x and y directions, which results in large divergence angle of the output beam. The simulated far-field emission pattern of the laser along the x and y directions are shown in Fig.~\ref{fig:QCL}(b).  The far-field profile is nearly symmetric with a full divergence angle of approximately 55$^\circ$. The proposed configuration for collimation of the QCL with the metasurface lens is shown schematically in Fig.~\ref{fig:QCL}(c). The metasurface lens has a focal length of 300 $\mu$m and its plane is parallel to the laser facet.

\begin{figure}
\centering
\includegraphics[width=0.8\columnwidth]{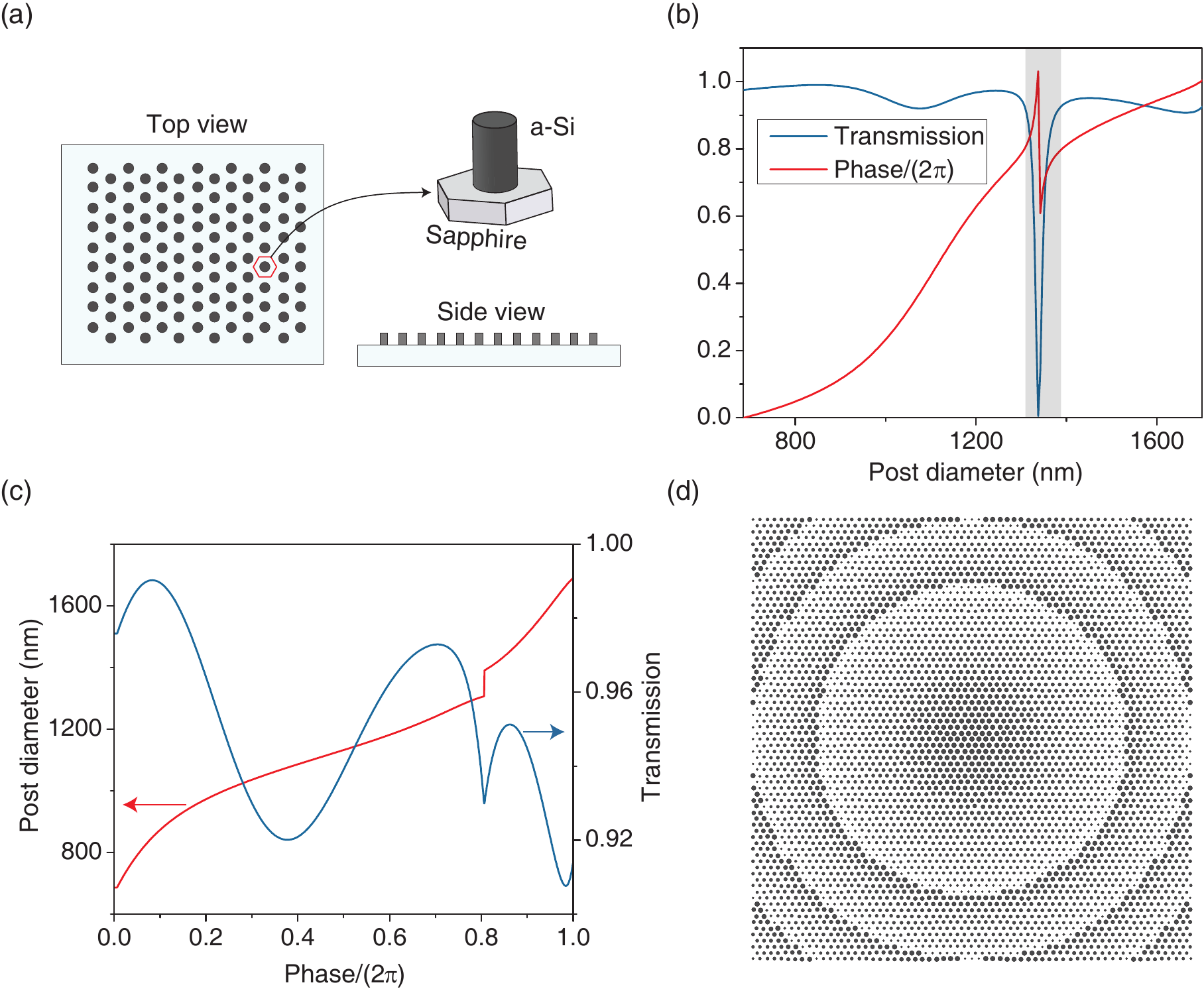}
\caption{(a) Schematic drawing of the top and side views of an array of circular posts which is used as a basis for implementation of the metasurface lens. The inset shows a unit cell of the array. (b) Intensity transmission coefficient and phase of the transmission coefficient for the array shown in (a) as functions of the post diameter. (c) One to one post diameter versus desired phase relation used in the design of the metasurface lens. The corresponding transmission values for different phases are also presented. (d) Schematic top view of the metasurface collimating lens.}
\label{fig:Lens_design}
\end{figure}

The lens is composed of an array of a-Si posts of different diameters which are arranged on a hexagonal lattice (inset of Fig.~\ref{fig:QCL}(c)). The posts are resting on a sapphire substrate. Each of the a-Si posts can be considered as a short waveguide with circular cross section truncated on both sides operating as a low-quality-factor Fabry-P\'{e}rot resonator. The circular cross section of the posts leads to the polarization insensitivity of the lens. Because of the high refractive index contrast between the posts and their surroundings, the posts behave as independent scatterers with small cross coupling~\cite{Arbabi2015}. The phase and amplitude of the scattered light depend on the diameter of the posts. It has been shown previously in the near-infrared region that the phase of the transmitted light, which is the sum of the incident and forward scattered light, can be controlled to take any value in the 0-2$\pi$ range by properly selecting the post diameter. The local transmission coefficient of an array of posts with gradually varying diameters can be approximated by the transmission coefficient of a uniform periodic array of posts. Figure~\ref{fig:Lens_design}(a) shows a uniform array of a-Si posts on a sapphire substrate whose diameter-dependent transmission coefficient is used to approximate the local transmission coefficient of the metasurface lens. The simulated intensity transmission coefficient, and the phase of the amplitude transmission coefficient for this array as a function of the posts diameter are shown in Fig.~\ref{fig:Lens_design}(b). The post height of 2.93 $\mu$m, lattice constant of 2.45 $\mu$m, and wavelength of 4.8 $\mu$m are assumed, and the simulation is performed by using the rigorous coupled wave analysis (RCWA) technique using a freely available software package~\cite{Liu2012}. The simulation assumes periodic boundary condition for the hexagonal unit cell shown as inset in Fig.~\ref{fig:Lens_design}(a). This reduces the computational resources required compared to simulation of scattering from a single post. The refractive index of sapphire was assumed as 1.63 and the refractive index of a-Si was determined to be 3.37 by extrapolating the refractive index obtained from spectroscopic ellipsometry measurements over the 0.4 $\mu$m to 2 $\mu$m range. The norm of the reciprocal lattice vectors of the array are smaller than the wavenumber in both sapphire and air; therefore, the array is non-diffracting for normal incidence. As Fig.~\ref{fig:Lens_design}(b) shows, the phase imparted by the posts varies by 2$\pi$ as the post diameter changes from 685 nm to 1700 nm. By excluding the post diameters corresponding to the reflective resonance (i.e. the transmission dip highlighted by a gray rectangle in Fig.~\ref{fig:Lens_design}(b)) from the design, the intensity transmission is kept above 91\%, and a one to one relationship between the phase and post diameter can be obtained as shown in Fig.~\ref{fig:Lens_design}(c). Note that the metasurface platform does not operate in the effective index regime~\cite{Chen1995}, and this enables realization of efficient high NA lenses~\cite{Lalanne1999}.

The subwavelength lattice constant and the large number of phase steps provided by the continuous post diameter-phase relation, enables accurate implementation of any exotic phase profile optimized for specific applications. To design a collimating lens for the QCL shown in Fig.~\ref{fig:QCL}(a), we first found the electrical field of the lasing mode through finite element simulation. As the simulated electric field of the lasing mode presented in the inset of Fig.~\ref{fig:QCL}(a) shows, the lasing mode is almost circular with subwavelength full width at half maximum spot size along both the x and y directions. Then, the electric field of the laser emission was found on the plane of the lens (i.e. a plane 300 $\mu$m away from the laser) through plane wave expansion technique~\cite{Born1999}, and by assuming that the electric field at the laser's cleaved facet has approximately the same distribution as the lasing mode's electric field. The simulated intensity and phase on this plane are shown as insets in  Fig.~\ref{fig:QCL}(c). The desired phase shift imparted by the collimating lens was set equal to the negative of the phase of the electric field of the laser at the lens plane such that the light transmitted through the lens has a flat wavefront. To implement a metasurface that imparts the desired spatially varying phase shift, the diameters of the posts at each location on the metasurface were obtained from Fig.~\ref{fig:Lens_design}(c) and the corresponding desired phase shift value at that location.  A schematic drawing of the top view of the metasurface lens is shown in Fig.~\ref{fig:Lens_design}(d). The focal length of the metasurface (which is also equal to its working distance) was set to 300~$\mu$m to facilitate the alignment of the lens to the QCL while keeping the lens aperture diameter reasonably small. We chose a lens diameter of 1 mm (corresponding to 0.86 NA). From the simulated intensity result shown in the inset of Fig.~\ref{fig:QCL}(c), we found that 87\% of the mid-IR light emitted by the laser passes through the lens aperture.

\begin{figure}
\centering
\includegraphics[width=0.8\columnwidth]{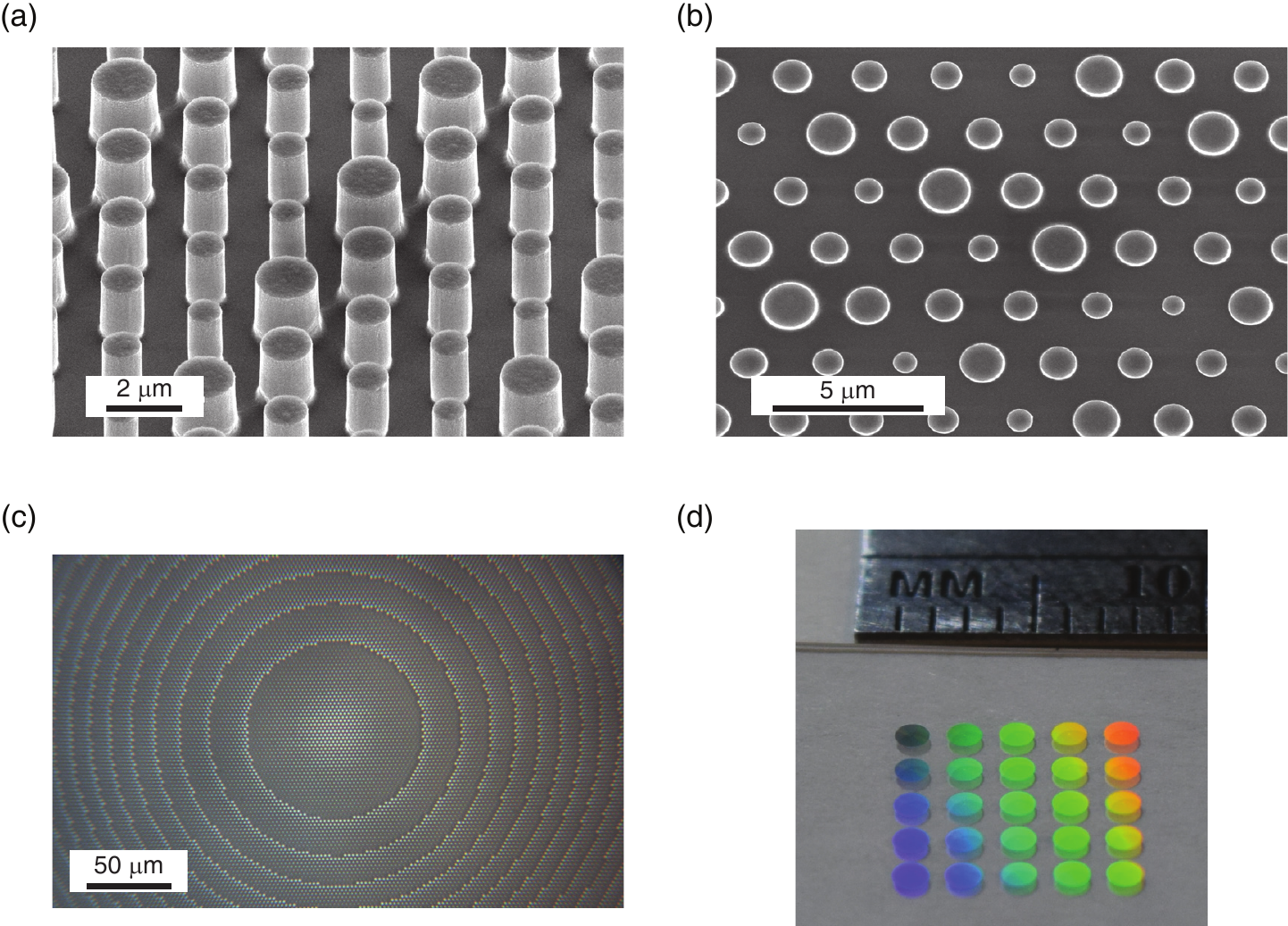}
\caption{(a) Scanning electron micrograph showing tilted and (b) top views of a fabricated metasurface lens. (c) Optical microscope image of the center part of a metasurface lens. (d) A 5$\times$5 array of metasurface lenses.}
\label{fig:fabricated_lenses}
\end{figure}

To fabricate the metasurface lenses, an approximately 3~$\mu$m thick layer of hydrogenated a-Si was deposited on a 430~$\mu$m thick double-side polished sapphire substrate. The a-Si was deposited by plasma enhanced chemical vapor deposition (PECVD) technique using a 5\% silane in argon mixture. The cross section of the sapphire wafer was imaged using scanning electron microscopy and the a-Si thickness was found to be 2.93~$\mu$m. Then, a 300 nm thick positive electron beam resist (ZEP-520A) was spun on the a-Si. To avoid charging effects, a charge dissipation polymer (Aquasave, Mitsubishi Rayon) was spin coated on the resist. The lens pattern was written on the resist using a 100 kV electron beam lithography system (EBPG-5000+, Leica Microsystems).  After removing the charge dissipation layer and developing the pattern, a 100 nm thick layer of aluminum oxide was deposited on the electron beam resist using electron beam evaporation. The aluminum oxide layer was pattered by lifting off the resist in a solvent (Remover PG, MicroChem Corporation) and was used as a hard mask in etching of the a-Si layer. The a-Si was etched using an inductively coupled plasma reactive ion etching (ICP-RIE) process with a mixture of \ce{SF_6} and \ce{C_4F_8} gases. The aluminum oxide mask was subsequently removed through wet etching in a 1:1 mixture of ammonium hydroxide and hydrogen peroxide heated to 80$^\circ$C. Note that the minimum feature size of the metasurface lenses is 685 nm and the patterning can also be done using widely available conventional UV lithography systems. Tilted and top views of the fabricated a-Si posts that compose the metasurface lenses are shown in Fig.~\ref{fig:fabricated_lenses}(a) and~\ref{fig:fabricated_lenses}(b), respectively. Figures ~\ref{fig:fabricated_lenses}(a) and~\ref{fig:fabricated_lenses}(b) show the posts at locations far from the lens center where phase varies rapidly.  An optical microscope image of the center part of a fabricated metasurface lens is shown in Fig.~\ref{fig:fabricated_lenses}(c). An array of 25 similar lenses was fabricated, as shown in Fig.~\ref{fig:fabricated_lenses}(d), each with a diameter of 1 mm.

\begin{figure}
\centering
\includegraphics[width=0.9\columnwidth]{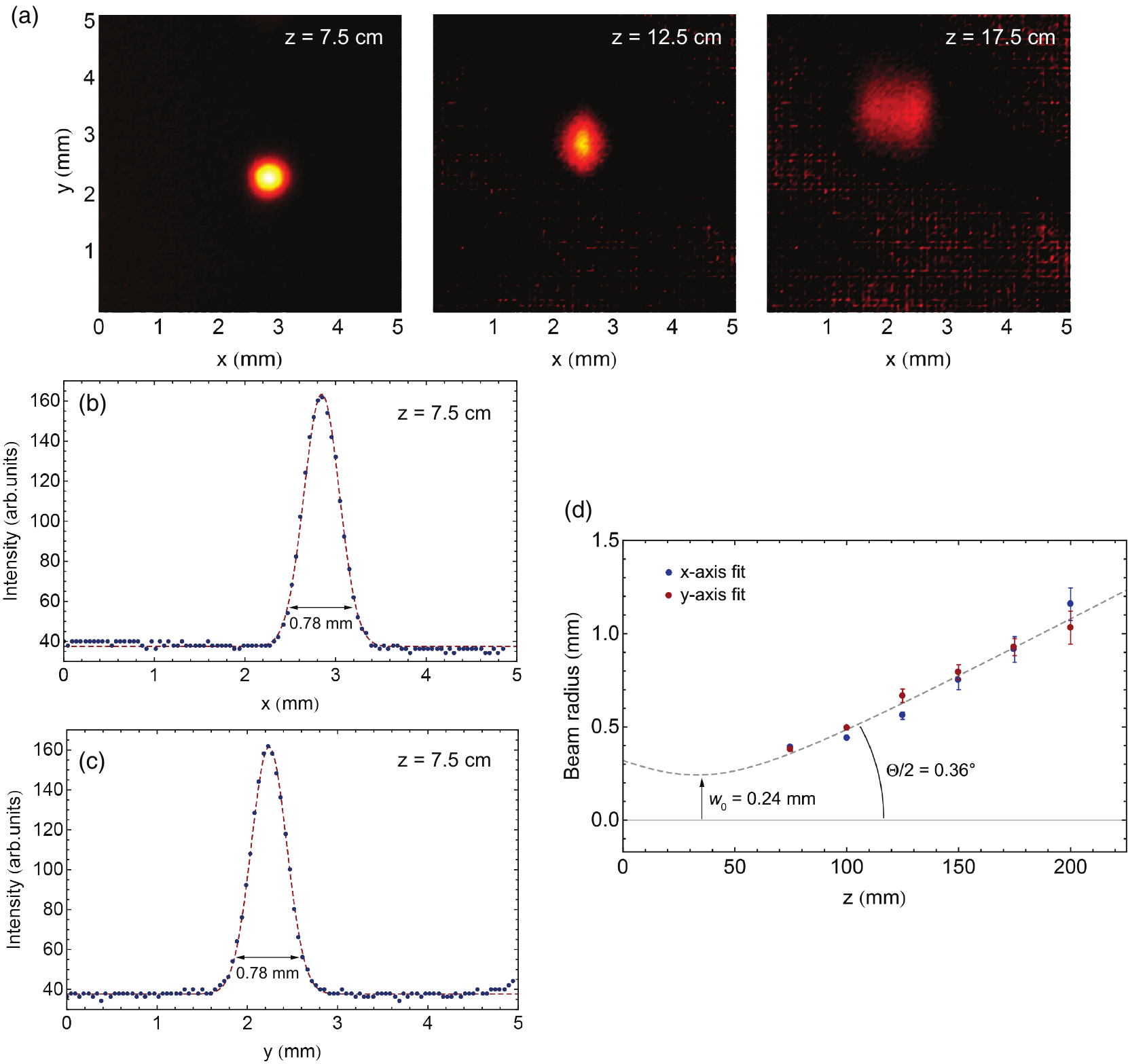}
\caption{(a) Measured transverse plane intensity profiles of the beam collimated by a metasurface lens at three different axial distances. (b) x and (c) y-cut intensity profiles and the corresponding best Gaussian fits at $z=7.5$ cm. $1/e^2$ beam diameters are also shown. (d) Measured beam radius of the collimated beam as a function of axial distance and best theoretical fit to the data.}
\label{fig:Characterization_results}
\end{figure}

The fabricated metasurface lenses were characterized using the DFB-QCL described previously. The laser was driven in continuous wave mode with an injection current of 130 mA and submount temperature of 30$^\circ$C. At these operating conditions, the output power was 11.4 mW and the emission wavelength was near 4.82~$\mu$m. More details regarding laser performance can be found in Ref.~\cite{Briggs2014a}. The array of metasurface lenses was mounted approximately 300 $\mu$m away from the laser facet (as schematically shown in Fig.~\ref{fig:QCL}c), and a lens was aligned to the laser to obtain a collimated output beam. The transverse intensity of the beam was recorded at different locations along the beam propagation axis (z axis) using an infrared camera (Electrophysics PV320). A neutral density filter was inserted before the camera to avoid saturation. Measured intensity profiles at different axial distances are shown in Fig.~\ref{fig:Characterization_results}(a). The non-zero background intensity observed in the recorded intensity profiles are due to thermal background and the camera's noise.  Gaussian functions were fit to the x and y-cuts of the measured intensity profiles (as shown in Fig.~\ref{fig:Characterization_results}(b) and~\ref{fig:Characterization_results}(c)) and the corresponding  beam radii were found on different z-planes separated from each other by steps of 2.5 cm. These z-dependent beam radii are shown in Fig.~\ref{fig:Characterization_results}(d) along with a theoretical fit for the z-dependent beam radius of a Gaussian beam (i.e. $w=w_0\sqrt{1+(z/z_0)^2}$  where $z_0=\pi w_0^2/\lambda$ is the Rayleigh range). Beam waist radius of $w_0=240~\mu$m and divergence half-angle of $\Theta/2=\lambda/(\pi w_0)=0.36^\circ$  were obtained from the fit. The beam radius of the laser beam at the location of the metasurface lens was estimated as $w_\mathrm{in}=247~\mu$m by fitting a two-dimensional Gaussian function to the simulated intensity result shown in the inset of Fig.~\ref{fig:QCL}(c); therefore, the beam quality factor was estimated as $M^2=w_{in}/w_0=1.02$. Note that $M^2$ values for collimated single mode diode lasers are typically in the range of 1.1 to 1.7~\cite{MGO_ptics_guide}. 

Using a calibrated thermopile detector, the output power of the DFB-QCL was measured to be 11.4 mW directly at the emission facet. With the metasurface lens in place and the detector placed 5 cm from the lens, the output power was measured to be 9.0 mW, corresponding to a total transmission efficiency of 79\%. The power meter had an input aperture diameter of 1.8 cm, and was zeroed to the background with the laser off. It should be noted that some of the power detected by the power meter might have passed through the lens without being diffracted and collimated by the lens. Based on the laser's far-field emission profile (shown in Fig.~\ref{fig:QCL}b), the distance between the laser and the power meter, and the power meter aperture size, we find that 8.2\% of the total laser power would reach the power meter when the lens is removed. This provides an upper limit for the possible non-diffracted optical power that might have been detected by the power meter; therefore, we estimate that the collimation efficiency is larger than 70.8\%.
Considering that 87\% of the QCL emission passes through the lens aperture, assuming 96\% transmission for the posts (average of the simulated transmission values presented in Fig.~\ref{fig:Lens_design}(c)), and taking into account $\sim$6\% reflection at the sapphire/air interface at the backside of the substrate, we obtain expected efficiency of $\sim$79\% which agrees well with the measured value.

In summary, we proposed and experimentally demonstrated efficient dielectric metasurface mid-IR lenses composed of a-Si posts on sapphire. These lenses are polarization insensitive and provide the high NA required for collimation of mid-IR QCLs. The flat form factor, low weight, freedom in the design of the lens phase profile, and potential for low cost wafer scale fabrication make these metasurface lenses attractive for incorporation in mid-IR devices and systems.    

\section*{Funding Information}
This work was supported by the Caltech/JPL President’s and Director’s Fund (PDF). R.M.B gratefully acknowledges support from the NASA PICASSO program. Y.H. was supported as part of the U.S. Department of Energy ``Light–Material Interactions in Energy Conversion" Energy Frontier Research Center under grant DE-SC0001293 and a Japan Student Services Organization (JASSO) fellowship. 

\section*{Acknowledgments}
The device fabrication was performed in the Kavli Nanoscience Institute at Caltech. The measurements were carried out at the Jet Propulsion Laboratory, California Institute of Technology, under contract with the National Aeronautics and Space Administration.

\clearpage

\end{document}